\documentclass[aps,pra,onecolumn,nofootinbib,superscriptaddress]{revtex4-2}

\usepackage{amsmath,amssymb,mathtools}
\usepackage{graphicx,braket}
\usepackage{hyperref}
\usepackage{bm,bbm,times}

\newtheorem{theorem}{Theorem}

\begin{document}
	
\title{Symmetric and Antisymmetric Quantum States from Graph Structure and Orientation}
	
\author{Matheus R. de Jesus}
\email{jesus1@ufpr.br}
\affiliation{Department of Physics, Federal University of Paran\'a, P.O. Box 19044, Curitiba 81531-980, Paran\'a, Brazil}
	
\author{Eduardo O. C. Hoefel}
\email{hoefel@ufpr.br}
\affiliation{Department of Mathematics, Federal University of Paran\'a, P.O. Box 19081, Curitiba 81531-980, Paran\'a, Brazil}
	
\author{Renato M. Angelo}
\email{renato.angelo@ufpr.br}
\affiliation{Department of Physics, Federal University of Paran\'a, P.O. Box 19044, Curitiba 81531-980, Paran\'a, Brazil}
	
\begin{abstract}
Graph states provide a powerful framework for describing multipartite entanglement in quantum information science. In their standard formulation, graph states are generated by controlled-$Z$ interactions and naturally encode symmetric exchange properties. Here we establish a precise correspondence between graph topology and exchange symmetry by proving that a graph state is fully symmetric under particle permutations if and only if the underlying graph is complete. We then introduce a generalized graph-based construction using a non-commutative two-qudit gate, denoted $GR$, which requires directed edges and an explicit vertex ordering. {We show that complete directed graphs generate fully antisymmetric multipartite states when endowed with appropriate orientations}. Together, these results provide a unified graph-theoretic description of bosonic and fermionic exchange symmetry based on graph completeness and edge orientation.
\end{abstract}
	
\maketitle

\section{Introduction}

In the early 2000s, a quantum formalism emerged that established a close connection between multipartite entangled states and graph theory. This formalism, together with its subsequent generalizations, underpins the implementation of measurement-based quantum computation (MBQC) \cite{Raussendorf2003,Raussendorf2001,walther2005experimental}
and has been extensively employed in a wide range of quantum information tasks \cite{Salem2024,PhysRevLett.95.120405,PhysRevLett.116.070401,houdayer2024solvable}. Known as graph states, this framework provides a transparent description of multipartite entanglement in terms of graph topology and has been comprehensively reviewed in Ref.~\cite{Hein2006}. An intriguing feature of graph states is the nontrivial relationship between the topological properties of the underlying graph and the exchange symmetry of the associated quantum state. In light of the growing number of recent experimental realizations of graph states \cite{Thomas2024,Cooper2024,Thomas2022,PRXQuantum.6.010305,Lee2023graphtheoretical} and their increasing relevance for quantum networks \cite{Fan2025,10835135}, it is timely to examine in detail how graph topology constrains the symmetry
properties of the corresponding quantum states.

{From a broader perspective, graph states form a central subclass of stabilizer states, in which structural properties of graphs translate into operational features of multipartite entanglement \cite{Gottesman1997,VandenNest2004,Schlingemann2002}. This correspondence has been extensively explored for undirected graphs and qubit systems, and admits natural extensions to higher-dimensional graph and cluster states, as well as to graph-based quantum codes \cite{Zhou2003,Looi2008,Ionicioiu2012,Helwig2013}. In parallel, permutation symmetry has long been recognized as a fundamental organizing principle for multipartite quantum correlations \cite{Markham2011}, while the fully antisymmetric sector (often discussed in terms of ``supersinglets'') constitutes a paradigmatic family of maximally entangled multipartite states \cite{Cabello2003}. Despite these parallel developments, the role of graph topology in enforcing exchange symmetry at the level of multipartite quantum states has not been \mbox{systematically characterized.}}

It has been noted in the literature that graph states associated with complete graphs exhibit permutation symmetry. Motivated by this observation, it is shown in this work that, within the standard graph-state formalism, a state is fully symmetric under particle permutations if and only if the graph from which it is constructed is complete. This characterization naturally raises a complementary question: whether an analogous graph-based construction exists for antisymmetric states. As demonstrated below, the conventional graph-state formalism does not allow for the generation of fully antisymmetric states, revealing a fundamental limitation of the standard approach.

To address this limitation, an extension of the graph-state formalism is introduced. The proposed framework relies on alternative two-qudit operators and directed graph structures, thereby enabling the systematic construction and analysis of antisymmetric multipartite states. Within this generalized setting, graph topology---and, crucially, edge orientation---plays a central role in determining the exchange symmetry of the resulting quantum states.

This article is organized as follows. After reviewing the necessary theoretical background on graph states and exchange symmetries, the equivalence between graph completeness and permutation symmetry is established, and the intrinsic obstruction to antisymmetry in the standard formalism is analyzed. {A generalized graph-based construction capable of generating antisymmetric states is then introduced and illustrated with explicit examples. Finally, the implications of these results are discussed, and possible directions for future research are outlined.}

\section{Exchange Symmetry and Generalized Graph States}

\subsection{Exchange Symmetry in Quantum Systems}

Exchange symmetry plays a fundamental role in the description of systems composed of identical quantum particles. Independently of the specific physical realization, the admissible states of such systems are constrained by their transformation properties under particle permutations. For a system of $N$ identical particles, any physical state must satisfy
\begin{align}
    &P_\sigma |\psi\rangle = |\psi\rangle
    &\quad &\text{for bosons,} \\
    &P_\sigma |\psi\rangle = \operatorname{sgn}(\sigma)|\psi\rangle
    &\quad &\text{for fermions,} \label{eq2}
\end{align}
where $P_\sigma$ denotes the unitary operator implementing the permutation
$\sigma \in S_N$ of the $N$ subsystems, and $\operatorname{sgn}(\sigma)$
is the signature (or parity) of the permutation, equal to $+1$ for even
permutations and $-1$ for odd ones. These relations provide the operational definition of bosonic and fermionic exchange symmetry employed throughout this work.

\subsection{Graph States via \boldmath{$CZ$}}

We now briefly recall the standard construction of graph states, which will serve as the reference framework for our subsequent analysis. Following Hein {\it et al.}~\cite{Hein2006}, the graph state associated with a simple undirected graph $G=(V,E)$ is defined as
\begin{equation}
    \label{eq:CZstate}
    |G\rangle = \prod_{(a,b)\in E} CZ^{(a,b)} |+\rangle^{\otimes N},
\end{equation}
where $V=\{1,\dots,N\}$ is the vertex set (one vertex per qubit) and $E \subseteq V\times V$ is the edge set specifying which pairs of qubits interact. Each vertex represents a qubit prepared in the state $|+\rangle = (|0\rangle + |1\rangle)/\sqrt{2}$, and each edge corresponds to a controlled-phase interaction of the Ising type between the associated qubits. Operationally, $CZ^{(a,b)}$ acts as a controlled phase flip, leaving all computational basis states unchanged except for $\ket{11}$, which acquires a phase factor $-1$. For example, applying $CZ^{(1,2)}$ to the product state $\ket{+}_1\otimes\ket{+}_2$ yields
$CZ^{(1,2)}\ket{+}_1\otimes\ket{+}_2=\frac{1}{2}(\ket{00}+\ket{01}+\ket{10}-\ket{11})$. Strictly speaking, the symbol $CZ^{(a,b)}$ is a compact notation for the operator
$CZ^{(a,b)} \otimes \bigotimes_{i\neq a,b}\mathbbm{1}_i$, meaning that the controlled-phase gate acts nontrivially only on the Hilbert space $\mathcal{H}_a\otimes\mathcal{H}_b$ associated with vertices $a$ and $b$, while acting as the identity on all remaining subsystems. As a result, the state $|G\rangle$ depends only on the topology of the underlying graph and not on the order in which the interactions are applied, since the operators $CZ^{(a,b)}$ mutually commute.

\subsection{Symmetry Properties of Graph States}

Let $\Gamma=(V,E)$ be a nontrivial simple graph, and let $\ket{G}$ denote the associated graph state defined in Equation \eqref{eq:CZstate}.

\begin{theorem}
{The graph state $\ket{G}$ is invariant under all permutations of the $N$ qubits if and only if the graph $\Gamma$ is complete.}
\end{theorem}

The ``if'' direction follows from the fact that, for the complete graph $K_N$, any permutation of vertices leaves the edge set invariant. (Here and in what follows, $K_N$ denotes the complete graph on $N$ vertices). Since all controlled-$Z$ gates commute, the action of a permutation operator $P_\sigma$ merely relabels the gates appearing in Equation \eqref{eq:CZstate}, leaving the state unchanged.

Conversely, we show that if $\Gamma$ is not complete, the corresponding graph state $\ket{G}$ fails to be permutation symmetric. Any non-complete graph necessarily contains at least one of the following minimal substructures that break symmetry:

\begin{itemize}
    \item[$h_1$:] Three vertices connected linearly, i.e., edges $(1,2)$ and $(2,3)$ are present, while $(1,3)$ is absent;
    \item[$h_2$:] Two disconnected vertices, at least one of which is connected to a third vertex.
\end{itemize}

The subgraph $h_1$ appears in every non-complete connected graph, while $h_2$ appears in every disconnected graph. Both structures can be obtained from the complete graph $K_3$ by the removal of one edge ($h_1$) or two edges ($h_2$), respectively. Before proceeding with the proof, we establish that these substructures necessarily occur in the corresponding classes of graphs.

\subsubsection{Structure $h_1$: Connected Graphs}

By definition, in a connected graph there exists a path between any two vertices. Among all such paths, consider one of minimal length. To show that the structure $h_1$ appears in every non-complete connected graph, we choose two non-adjacent vertices $u$ and $v$ of $\Gamma$. Such a choice is always possible since the graph is not complete.

Let $(u = w_0, w_1, \ldots, w_{k}, w_{k+1} = v)$ be a shortest path connecting $u$ and $v$, where $k \geq 1$ is the number of intermediate vertices.

\begin{itemize}
\item If $k=1$, the vertices $(u, w_1, v)$ form exactly the structure $h_1$, with edges $(u,w_1)$ and $(w_1,v)$ and no edge $(u,v)$.
\item If $k>1$, any three consecutive vertices $(w_i, w_{i+1}, w_{i+2})$ along the path form a connected subgraph isomorphic to $h_1$.
\end{itemize}

Therefore, every non-complete connected graph necessarily contains the substructure $h_1$.

\subsubsection{Structure $h_2$: Disconnected Graphs}

In a disconnected graph, there exists no path between vertices belonging to different connected components. For a disconnected graph to be nontrivial, it must contain at least one edge. Therefore, one can always choose two non-adjacent vertices such that one of them is adjacent to a third vertex, yielding a subgraph isomorphic to the structure $h_2$.

\subsubsection{Symmetry Breaking in the Minimal Substructures}

Consider first the connected case $h_1$.
For vertices $1$, $2$, and $3$, the corresponding graph state is
\begin{align}
\ket{G}_{h_1}
&= CZ^{(1,2)}\, CZ^{(2,3)} \ket{+}^{\otimes 3} \notag \\
&= \frac{1}{\sqrt{8}} \Big(
\ket{000} + \ket{100} + \ket{010} + \ket{001}
- \ket{110} + \ket{101} - \ket{011} + \ket{111}\Big).
\end{align}
Applying the permutation $P_{12}$, we obtain
\begin{equation}
P_{12}\ket{G}_{h_1}
= \frac{1}{\sqrt{8}} \Big(
\ket{000} + \ket{010} + \ket{100} + \ket{001}
- \ket{110} + \ket{011} - \ket{101} + \ket{111}
\Big),
\label{eq5}
\end{equation}
which is clearly different from the original state.
Hence, $P_{12}\ket{G}_{h_1} \neq \ket{G}_{h_1}$, showing that permutation symmetry is broken.

For the disconnected case $h_2$, where only one edge connects vertices $(1,2)$, the associated state reads
\begin{equation}
\ket{G}_{h_2}
= CZ^{(1,2)} \ket{+}^{\otimes 3}
= \frac{1}{\sqrt{8}} \Big(
\ket{000} + \ket{100} + \ket{010} + \ket{001}
- \ket{110} + \ket{101} + \ket{011} - \ket{111}
\Big).
\end{equation}
Under the permutation $P_{23}$, we find
\begin{equation}
P_{23}\ket{G}_{h_2}
= \frac{1}{\sqrt{8}} \Big(
\ket{000} + \ket{100} + \ket{001} + \ket{010}
- \ket{101} + \ket{110} + \ket{011} - \ket{111}\Big),
\end{equation}
again yielding $P_{23}\ket{G}_{h_2} \neq \ket{G}_{h_2}$.

To elucidate why the presence of such a substructure breaks the symmetry of a larger graph, consider an explicit example in which a fourth vertex is added to the structure $h_1$. The resulting state can be written as
\begin{equation}
\prod_{(i,j)\in E_G \setminus E_{h_1}} CZ^{(i,j)}
\big[ \ket{G}_{h_1} \otimes \ket{+} \big].
\end{equation}
For concreteness, we choose the set of additional edges
$E_G \setminus E_{h_1} = \{(1,4),(2,4),(3,4)\}$.
This yields
\begin{align}
CZ^{(1,4)} CZ^{(2,4)} CZ^{(3,4)}
&\big[ \ket{G}_{h_1} \otimes \ket{+} \big]
= \frac{1}{4} CZ^{(1,4)} CZ^{(2,4)} CZ^{(3,4)}
\Big(\ket{000} + \ket{100} + \ket{010} \notag \\
&+ \ket{001} - \ket{110} + \ket{101} - \ket{011} + \ket{111}
\Big) \otimes (\ket{0} + \ket{1}).
\end{align}

Recall that the action of the gate $CZ^{(i,4)}$ is to apply a phase factor $-1$ whenever both qubits $i$ and $4$ are in the state $\ket{1}$. Consequently, the controlled-$Z$ operations affect only the component of the state proportional to $\ket{1}_4$.
The state can therefore be rewritten as 
\begin{equation}
\frac{1}{4}\Big[
\ket{G}_{h_1} \otimes \ket{0} + \big(
\ket{000} - \ket{100} - \ket{010} - \ket{001}
- \ket{110} + \ket{101} - \ket{011} - \ket{111}
\big) \otimes \ket{1} \Big].
\end{equation}

Finally, applying the same permutation $P_{12}$ as in Equation \eqref{eq5}, it follows immediately that the total state is not symmetric, since
$P_{12}\ket{G}_{h_1} \neq \ket{G}_{h_1}$.
This construction can be generalized to an arbitrary graph $\Gamma$ with $N$ vertices:
\begin{align}
\ket{G}_{\Gamma}
&= \prod_{(i,j)\in E_\Gamma \setminus E_{h_1}}
CZ^{(i,j)}\, \ket{G}_{h_1} \otimes \ket{+}^{\otimes (N-3)} \notag \\
&= \frac{1}{2^{(N-3)/2}}
\prod_{(i,j)\in E_\Gamma \setminus E_{h_1}}
CZ^{(i,j)}\, \ket{G}_{h_1} \otimes
\big( \ket{0} + \ket{1} \big)^{\otimes (N-3)} .
\label{eqProva}
\end{align}

By the same reasoning, the state defined in Equation \eqref{eqProva} is not permutation-symmetric, since the nonsymmetric substructure $\ket{G}_{h_1}$ appears in a tensor product with $\big( \ket{0} + \ket{1} \big)^{\otimes (N-3)}$. Consequently, the component $\ket{G}_{h_1}\otimes\ket{00\ldots0}$ remains unaffected by the additional commuting $CZ$ gates. The choice of the subgraph $h_1$ was made solely for illustrative purposes, and the argument applies equally to the structure $h_2$. Therefore, any non-complete graph contains at least one of these substructures and thus gives rise to a non-symmetric graph state. The only topology whose associated state remains invariant under all permutations is the complete graph $K_N$.

This result establishes a direct correspondence between the topological completeness of the graph and the full permutation symmetry of the associated graph state. It provides a simple but powerful criterion: only complete graphs generate symmetric graph states within the standard $CZ$ formalism.

\subsection{The Impossibility of Antisymmetry in Standard Graph States}

Within the standard $CZ$-based graph-state formalism, it is not possible to construct fully antisymmetric states. Let $\Gamma=(V,E)$ be a graph defining an $N$-qubit graph state $\ket{G}$ via Equation \eqref{eq:CZstate}. As discussed previously, the controlled-$Z$ gate $CZ^{(a,b)}$ acts on a pair of qubits $(a,b)$ as a controlled phase flip: whenever both qubits are in the state $\ket{1}$, the gate applies a phase factor $-1$.

Crucially, $CZ^{(a,b)}$ is diagonal in the computational basis. As a consequence, its action only modifies relative phases and does not change which computational-basis vectors appear with nonzero amplitude in the state. Therefore, the support of $\ket{G}$ coincides with that of the initial product state $\ket{+}^{\otimes N}$, which can be written explicitly as
\begin{equation}
\label{invariante}
\ket{+}^{\otimes N}= \frac{1}{2^{N/2}}
\big(\ket{00\cdots 0} + \ket{01\cdots 0} + \cdots + \ket{11\cdots 1}
\big).
\end{equation}

This state contains all $2^N$ computational-basis vectors with equal magnitude. Moreover, some of these components are invariant under both the action of all $CZ^{(a,b)}$ gates and any permutation of subsystems, such as the vector $\ket{00\cdots 0}$. Importantly, the amplitude of this component is nonzero and remains unchanged throughout the graph-\mbox{state construction.} 

Consequently, although certain basis vectors may acquire a minus sign under a given permutation, it is impossible for the total state $\ket{G}$ to satisfy the fermionic antisymmetry condition $P_\sigma\ket{G}=\operatorname{sgn}(\sigma)\ket{G}$ for all $\sigma\in S_N$, as required by Equation \eqref{eq2}. This establishes that fully antisymmetric states cannot arise within the standard graph-state formalism.

\subsection{The \boldmath{$GR$} Gate and Antisymmetric Graph States}

The impossibility of generating fully antisymmetric states within the standard $CZ$-based graph-state formalism naturally raises the question of whether an alternative graph-based construction can accommodate fermionic exchange symmetry. In particular, one may ask whether the obstruction identified above is intrinsic to graph-theoretic descriptions or rather a consequence of the specific choice of two-qubit interaction underlying conventional graph states.

To address this question, we consider a generalized framework based on a different class of two-body gates. While no graph-theoretic construction of antisymmetric states appears to exist in the literature, it is worth noting that, outside the context of graph states, antisymmetric multipartite states can be generated using quantum gates. For instance, Ref.~\cite{Jex2003} presents a construction of the fully antisymmetric state for three-dimensional systems using two-qubit gates. Generalizing such a construction to any dimension is one of the motivations for the present paper.

Motivated by these observations, we introduce here a gate-based formulation that enables the systematic construction of antisymmetric states and admits a natural interpretation in terms of directed graphs. The key ingredient is an asymmetric two-qudit gate, denoted $GR^{(l,k)}$~\cite{Jex2003}, acting on pairs of $n$-level systems (qudits). This operator is closely related to the generalized quantum XOR (GXOR) gate introduced in \cite{Alber2000}, differing only by a relabeling of the control and target registers and by modular inversion. Gates of this type are standard entangling operations in qudit quantum information and are closely related to the generalized controlled-addition (or SUM) gate widely used in qudit quantum circuits; see, e.g., \cite{Zhou2003,Wang2003}. The asymmetry of the $GR$ gate under index exchange will play a crucial role in the construction below, as it allows the orientation of graph edges to influence the symmetry properties of the resulting state. Its action on computational-basis states is defined by
\begin{equation}
GR^{(l,k)} \ket{i}_k \ket{j}_l
= \ket{j \ominus i}_k \ket{j}_l,
\end{equation}
where $\ominus$ denotes subtraction modulo $n$. Throughout this construction we take the local dimension $d$ of each qudit to be equal to the number of particles, i.e., $d=n$. This choice ensures that the computational basis $\{\ket{0},\ket{1},\ldots,\ket{n-1}\}$ provides enough distinct labels to represent all permutations appearing in the totally antisymmetric state. Note that for a fixed first index, $[GR^{(l,k)},GR^{(l,j)}]=0$ for all $j,k \in \mathbb{N}$. This commutativity can be verified directly from the definition of the $GR$ operator.

In addition, we introduce {a phase-adapted Hadamard operator} and shift operators:
\begin{align}
H_n \ket{0} &= \frac{1}{\sqrt{n}} \sum_{k=0}^{n-1} (-1)^{ k(n-1)}\ket{k}, \label{had} \\
X_n \ket{k} &= \ket{k+1 \bmod n}. \label{xd}
\end{align}
The operator $H_n$ is a phase-adapted version of the standard qudit Hadamard (or discrete Fourier transform) widely used in qudit quantum information; see, e.g., \cite{Wang2003}. The specific phase choice in \eqref{had} is tailored to the recursive construction introduced below. Note that when $n$ is odd, the relative phase factor becomes trivial.

Using these operators, fully antisymmetric states can be constructed recursively. Starting from the two-qudit antisymmetric Bell state
\[\ket{A_2} = \frac{1}{\sqrt{2}} \bigl(\ket{01} - \ket{10}\bigr),\]
we define the $n$-partite state $\ket{A_n}$ as
\begin{equation}
\ket{A_n}
= \prod_{i=1}^{n-1} GR^{(n,i)}
\left(
X_n^{\otimes (n-1)} \ket{A_{n-1}} \otimes H_n \ket{0}_n
\right).
\label{antis}
\end{equation}

As will be shown in the following subsections, this recursive construction admits a natural interpretation in terms of directed graphs, in which adjacency relations and edge orientation play a central role in determining the exchange symmetry of the resulting quantum states.

\subsubsection*{Full Antisymmetry of the State for an Arbitrary Number of Qudits}\label{subsec:prova}

The goal of this subsection is to show that the state $|A_n\rangle$ defined in Equation \eqref{antis} coincides, up to a global phase, with the fully antisymmetric state on $n$. The proof proceeds by rewriting $|A_n\rangle$ as a uniform superposition of computational-basis states labeled by permutations, with coefficients determined by their signatures. The key step is to show that the relative signs induced by the recursive construction are independent of the auxiliary summation index, allowing the sum to be identified with the alternator over $S_n$. This subsection is the only part of the text where some familiarity with permutation groups is required; we refer the reader to \cite{sagan2001symmetric}. 

The proof starts by expanding the formula \eqref{antis} using the definitions \eqref{had}, \eqref{xd} and the hypothesis that the state $\ket{A_{n-1}}$ is fully antisymmetric. This hypothesis is completely reasonable since the recursive definition of the $\ket{A_n}$, which is valid for $n>2$, starts with the well-known $\ket{A_2}$. Assuming that $\ket{A_{n-1}}$ is antisymmetric and using the shift operator, one has
\begin{equation}
    X_n^{\otimes (n-1)}\ket{A_{n-1}}=\sum_{\sigma\in S_{n-1}}\frac{1}{\sqrt{(n-1)!}}(-1)^{|\sigma|}\ket{\sigma_0+1}_1\ldots\ket{\sigma_{n-2}+1}_{n-1}.
    \label{anm1}
\end{equation}
Applying Equation \eqref{anm1} in \eqref{antis} and expanding the adapted Hadamard, one gets
\begin{equation}
     \ket{A_n}= \sum_{k=0}^{n-1}\sum_{\sigma\in S_{n-1}} \Big(\prod_{i=1}^{n-1}GR^{(n,i)}\Big)\frac{(-1)^{k(n-1)}}{\sqrt{n!}}\big[  (-1)^{|\sigma|}\ket{\sigma_{0}+1}_1\ldots\ket{\sigma_{n-2}+1}_{n-1}\otimes\ket{k}_n\big].
     \label{prov}
\end{equation}

By applying the $GR^{(n,i)}$ operators in \eqref{prov}, one obtains the following combinatorial description of the state $|A_n\rangle$:
\begin{equation}
    |A_n\rangle = \frac{1}{\sqrt{n!}} 
    \sum_{k = 0}^{n-1} \;
    \sum_{\sigma \in S_{n-1}}
    \hspace{-1ex} (-1)^{|\sigma|}(-1)^{k(n-1)}|k - (\sigma_0+1)\rangle_1 \; \cdots \;  |k - (\sigma_{n-2}+1)\rangle_{n-1} |k\rangle_n
    \label{an_comb}
\end{equation}
where $\sigma$ runs over the set of all permutations of the set $\{0, 1, \dots, n-2\}$ of $n-1$ elements. Hence, $1 \leqslant \sigma_i + 1 \leqslant n - 1$, for all $i = 0, \dots, n-2$. All the additions and subtractions in the above formula must be considered modulo $n$.

Given a permutation $\sigma : \{0, 1, \dots, n-2\} \to \{0, 1, \dots, n-2\}$, one can obviously extend it to a permutation $\sigma + {\rm Id}$ of $\{0, 1, \dots, n-2, n-1\}$ defined by 
$(\sigma + {\rm Id})_i = \sigma_i$ for $i \leqslant n-2$, and $(\sigma + {\rm Id})_{n - 1} = n-1$, i.e.,\ as the identity on $n-1$. The new permutation obtained by this simple extension has the same sign as $\sigma$, i.e., ${\rm sgn}(\sigma + {\rm Id}) = {\rm sgn}(\sigma)$. Notice now that the cyclic permutation 
$(0 \; 1 \; 2 \; \cdots \; n - 1)$ corresponds precisely to addition of $1$ modulo $n$, whose sign is $(-1)^{(n-1)}$. Hence, the composition 
$(0 \; 1 \; 2 \; \cdots \; n - 1) \circ (\sigma + {\rm Id})$ gives a permutation such that $i \mapsto \sigma_i + 1$, for $i \leqslant n-2$, and $n-1 \mapsto 0$. Recall that $n \equiv 0 \pmod n$. Composing with the additive inverse, i.e., adding the minus sign, changes the permutation (and hence its sign) in a way that is independent of $k$. The new permutation thus obtained can be described by $i \mapsto -(\sigma_i + 1)$, for $i \leqslant n-2$, and $n-1 \mapsto 0$. Its sign does not depend on $k$. 

By taking this last permutation and iteratively composing $k$ times with the cyclic permutation $(0 \; 1 \; 2 \; \cdots \; n - 1)$, one finally gets the permutation $\pi_{\sigma, k} : \{0, 1, \dots, n-1\} \to \{0, 1, \dots, n-1\}$, defined by 
$i \mapsto k - (\sigma_i + 1)$ for $i \leqslant n-2$, and $n-1 \mapsto k$. The signature of the cyclic permutation $(0\;1\dots n-1)$ is $(-1)^{n-1}$. Thus, composing it $k$ times yields $(-1)^{|\pi_{\sigma,k}|} = \pm(-1)^{|\sigma|}(-1)^{k(n-1)}$, and therefore $(-1)^{|\sigma|} = \pm(-1)^{|\pi_{\sigma,k}|}(-1)^{k(n-1)}$. (This holds because the signature of a composition of permutations is the product of their respective signatures.) By substituting all that into formula (\ref{an_comb}), one gets
\begin{equation}
	|A_n\rangle = \frac{1}{\sqrt{n!}} 
	\sum_{k = 0}^{n-1} \;
	\sum_{\sigma \in S_{n-1}}
	\hspace{-1ex} \big[\pm (-1)^{|\pi_{\sigma, k}|}(-1)^{k(n-1)}\big]\times(-1)^{k(n-1)}
	|\pi_{\sigma, k}(0)\rangle_1 \; \cdots \; |\pi_{\sigma, k}(n-2)\rangle_{n-1} |\pi_{\sigma, k}(n-1)\rangle_n.
	\label{an_comb_pik}
\end{equation}

Recall that $\pi_{\sigma,k}(i)=k-(\sigma_i+1)$. Observe that the map $(\sigma,k)\mapsto\pi_{\sigma,k}$ associates to each pair $(\sigma,k)\in S_{n-1}\times\mathbb{Z}_n$ a permutation of $\{0,\ldots,n-1\}$. At this step one should note the fact that these permutations run through all the permutations. To be precise, the composition of these operations defines a group; hence $k-g=k$ if and only if $g=0$, since $0$ is the identity of the additive group. As previously explained, $-(\sigma_i +1)\neq0$ for $i<n-1$ and in the $n-1$ position $\pi_{\sigma,k}(n-1)=k-(\sigma_{n-1}+1)=k$. As an illustration, let us consider one state of the sum, assuming that $\sigma'\in S_{n-1}$ is an arbitrary permutation:
\[
\ket{k-(\sigma'_0+1)}_1\otimes\ket{k-(\sigma'_1+1)}_2\otimes \cdots\otimes\ket{k}_{n}.
\]
Since this composition is the operation of a group, all terms in the state above are distinct. Since the sums in Equation \eqref{an_comb_pik} are taken over all $k$ and $\sigma$, it follows that the resulting state runs through the entire permutation group $S_n$.

Expanding the signs in Equation \eqref{an_comb_pik} one finally gets
\begin{align}
     |A_n\rangle = \pm\frac{1}{\sqrt{n!}} 
    \sum_{k = 0}^{n-1} \;
    \sum_{\sigma \in S_{n-1}}
    \hspace{-1ex} (-1)^{|\pi_{\sigma, k}|}
    |\pi_{\sigma, k}(0)\rangle_1 \; \cdots \; |\pi_{\sigma, k}(n-2)\rangle_{n-1} |\pi_{\sigma, k}(n-1)\rangle_n.
    \label{an_comb_f}
\end{align}
Since the permutations $\pi_{\sigma,k}$ run through the entire symmetric group $S_n$, it follows that $|A_n\rangle$ is a multiple of the alternator of $|012 \cdots n-1\rangle$. Given that the alternator is defined as the sum over all permutations with a sign determined by their parity, i.e.,  $\operatorname{Alt}(|\psi\rangle) = \sum_{\sigma \in S_n} \operatorname{sgn}(\sigma)\,|\sigma_{0}\sigma_{1}\ldots\sigma_{n-1}\rangle$, we have
\begin{equation}
|A_n\rangle = \frac{\pm 1}{\sqrt{n!}}\,{\rm Alt}(|012 \cdots n-1\rangle).
\end{equation}
This shows that $|A_n\rangle$ is fully antisymmetric. This result shows that the recursive $GR$-based construction yields the unique fully antisymmetric state on $n$ qudits, anticipating its interpretation as the graph state associated with a complete directed graph.

\subsection{Definition of Operators}

In order to make precise the operator-based formulation underlying the $GR$ construction and its graph-theoretic interpretation, we now refine the notation introduced in the recursive definition of antisymmetric states.

We begin with the shift operator $X_n$. In Equation \eqref{antis}, the operator
\begin{equation}
X_n^{\otimes (n-1)} = X_n^{(1)} \otimes X_n^{(2)} \otimes \cdots \otimes X_n^{(n-1)}
\end{equation}
acts uniformly on all subsystems labeled from $1$ to $n-1$. Here, the superscript $(i)$ explicitly indicates the subsystem on which the operator acts.

In graph states generated by the $GR$ operator, the uniform application of $X_n$ used in Equation \eqref{antis} is no longer appropriate, as the displacement must reflect the adjacency relations of the underlying directed graph. Accordingly, we introduce the modified operator
\begin{equation}
\tilde{X}_n = \bigotimes_{i \in N_n} X_n^{(i)},
\end{equation}
where $N_n$ denotes the set of vertices adjacent to vertex $n$.
This definition ensures that only subsystems directly connected to the newly added vertex undergo a modular displacement, thereby encoding the graph structure at the operator level.

We now turn to the two-qudit gate $GR$. In contrast with the controlled-$Z$ gate, which satisfies the symmetry $CZ^{(i,j)} = CZ^{(j,i)}$ and therefore does not require an orientation of the edges, the action of the $GR$ gate is order-sensitive. In general, $GR^{(i,j)} \neq GR^{(j,i)}$, so the orientation of each edge must be specified and must influence the operator’s action.

Let $\Gamma$ be a directed graph and let $e=(u,v)$ be an edge of $\Gamma$. We define the \mbox{oriented operator}
\begin{equation}
\widetilde{GR}^{(u,v)} =
\begin{cases}
GR^{(u,v)}, & \text{if } o(e)=u, \\
GR^{(v,u)}, & \text{if } o(e)=v,
\end{cases}
\end{equation}
where $o:E\to V$ is the function that assigns to each directed edge its origin vertex. From this point on, for notational simplicity, we omit the tilde and write $GR^{(u,v)}$, with the understanding that the operator ordering is determined by the orientation of the corresponding edge.

These definitions make explicit how adjacency relations and edge orientation are encoded at the operator level, thereby establishing the link between directed graph structure and the generation of antisymmetric multipartite states.

\subsection{Graph States Generated by the \boldmath{$GR$} Operator}

To construct the quantum state associated with a directed graph $\Gamma$, it is necessary to take into account not only its adjacency relations, but also its orientation and an explicit ordering of the vertices. The latter is essential, since the state associated with an arbitrary graph is obtained through a recursive procedure in which vertices are incorporated sequentially. 

We begin with the initial subgraph $\Gamma_2$ consisting of the first two vertices. If no edge exists between vertices $1$ and $2$, we define the corresponding state simply as
\begin{equation}
\ket{\Gamma_2} = \ket{0}_1 \otimes \ket{0}_2 .
\end{equation}

If, on the other hand, an edge is present between vertices $1$ and $2$ (in either orientation), the preparation of the initial state must reflect this adjacency. In this case, vertex $1$ is prepared in the state $X\ket{0}=\ket{1}$, while vertex $2$ is prepared in the state $-H(X\ket{0})=\frac{1}{\sqrt{2}}(\ket{1}-\ket{0})$. The initial two-vertex state is then given by
\begin{equation}
\ket{\Gamma_2}
= GR^{(i,j)}\!\left[
\ket{1} \otimes \frac{1}{\sqrt{2}}(\ket{1}-\ket{0})\right],
\end{equation}
where $(i,j)=(1,2)$ or $(i,j)=(2,1)$, depending on the orientation of the edge.

Starting from this initial condition, the state associated with the subgraph $\Gamma_n$ on $n$ vertices is constructed recursively. Whenever vertex $n$ has at least one neighbor, we define
\begin{equation}
\ket{\Gamma_n}
=\prod_{(n,j)\in E(\Gamma)}GR^{(n,j)}
\Big(\tilde{X}_n \ket{\Gamma_{n-1}}\otimes H_n \ket{0}_n\Big),
\label{Eqjesus2}
\end{equation}
where the product runs over all edges connecting vertex $n$ to previously introduced vertices, and $\tilde{X}_n$ is the modified shift operator defined in the previous subsection. In Equation \eqref{Eqjesus2}, the index $n$ is fixed, and $j$ runs over the set of vertices adjacent to $n$ in increasing order. As previously shown, the action of the $GR$ gate depends on the orientation of the corresponding edge. If vertex $n$ has no neighbors, no interaction is applied and the state is defined by
\begin{equation}
\ket{\Gamma_n} = \ket{\Gamma_{n-1}} \otimes \ket{0}_n .
\end{equation}
When the graph $\Gamma$ is complete {and the orientation $(i,j)$ of all edges respects the relation $i>j$}, one recovers the state $|A_n\rangle$; in other words, 
\[ |\Gamma_n\rangle = |A_n\rangle.\]

\subsubsection*{Illustrative Example}

\begin{figure}[htb]
\centering
\includegraphics[width=0.42\linewidth,trim=6 6 6 6,clip]{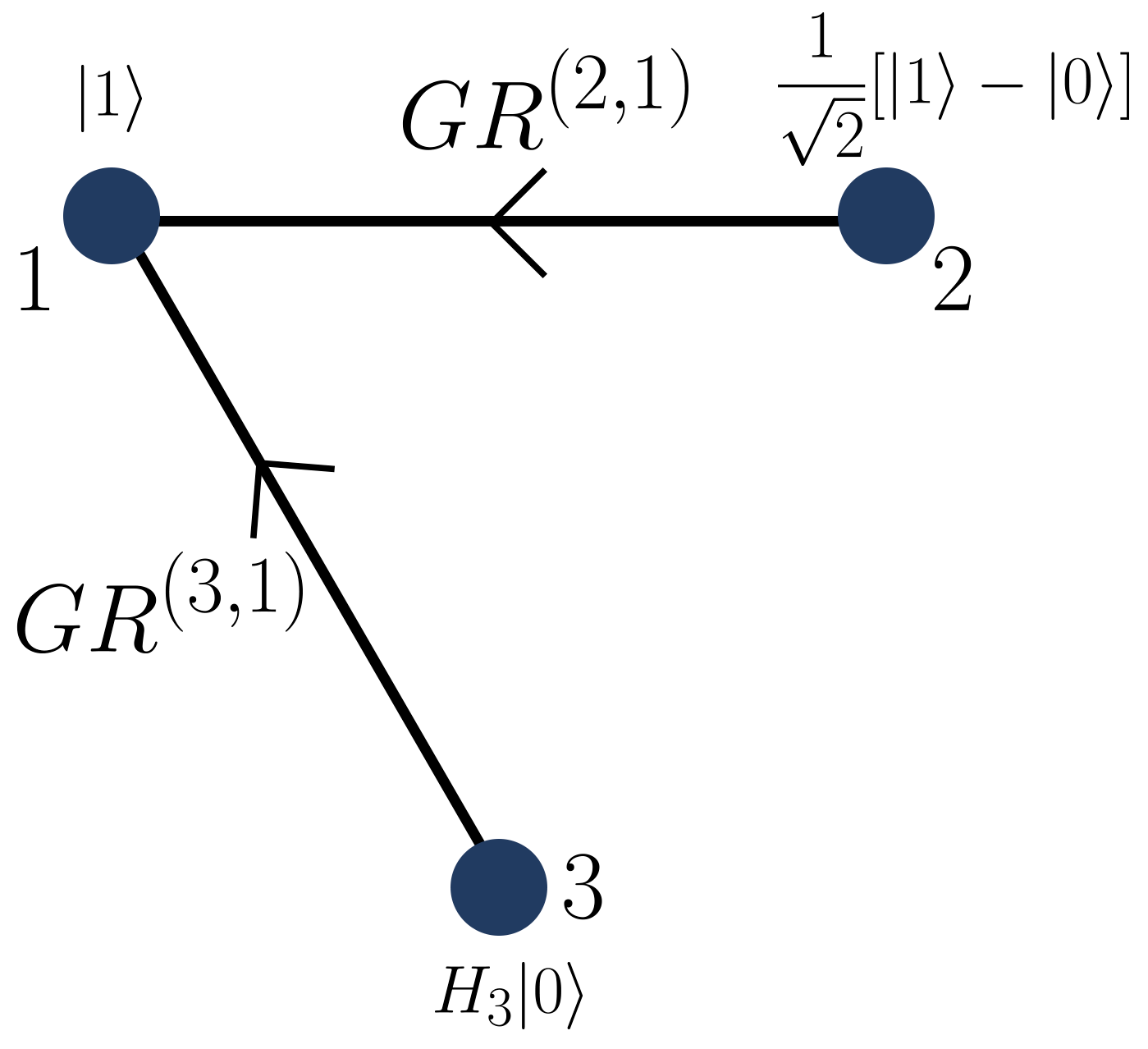}
\caption{Directed graph illustrating the intermediate step of the recursive construction of $\ket{\Gamma_3}$. The state $\ket{\Gamma_2}$ associated with vertices $1$ and $2$ is extended by adding vertex $3$, initially prepared in the state $H_3\ket{0}_3$. The directed edge $3\to1$ represents the application of the gate $GR^{(3,1)}$. At this stage, the resulting three-qudit state does not exhibit any particular exchange symmetry.}
\label{grafo4}
\end{figure}

\begin{figure}[htb]
\centering
\includegraphics[width=0.42\linewidth,trim=6 6 6 6,clip]{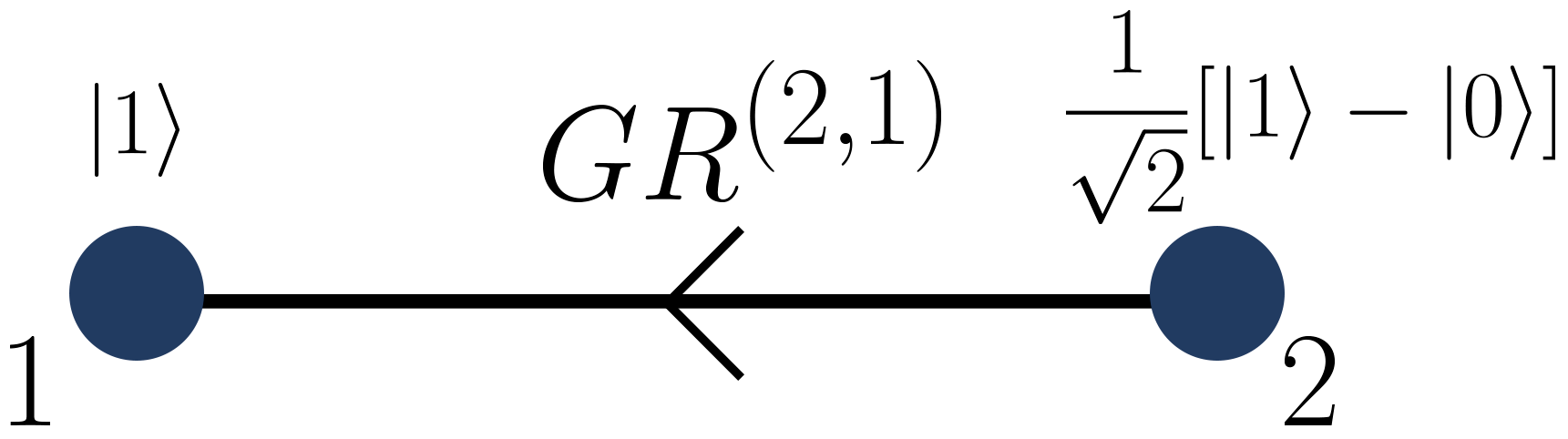} 
\caption{Initial two-vertex directed subgraph $\Gamma_2$ used as the base case of the recursive construction. Vertex $1$ is prepared in the state $\ket{1}$, while vertex $2$ is prepared in the state $\frac{1}{\sqrt{2}}(\ket{1}-\ket{0})=-H(X\ket{0})$. The directed edge from vertex $2$ to vertex $1$ encodes the application of the two-qudit gate $GR^{(2,1)}$, resulting in the antisymmetric state $\ket{\Gamma_2}=\frac{1}{\sqrt{2}}(\ket{01}-\ket{10})$.}
\label{grafo3}
\end{figure}

To illustrate the construction, we consider the directed graph shown in Figure~\ref{grafo4}. The corresponding initial two-vertex subgraph $\Gamma_2$, depicted in Figure~\ref{grafo3}, serves as the base case of the recursive procedure.\vspace{-4pt}


For the initial graph $\Gamma_2$, we obtain
\begin{equation}
\ket{\Gamma_2}=GR^{(2,1)}\!\left[
X_2\ket{0}\otimes\frac{1}{\sqrt{2}}(\ket{1}-\ket{0})\right]
=\frac{1}{\sqrt{2}}(\ket{01}-\ket{10}),
\end{equation}
which is already antisymmetric under exchange of the two subsystems.

Applying the recursive construction given in Equation \eqref{Eqjesus2}, and considering only the edge $3\to1$, we obtain
\begin{align}
\ket{\Gamma_3}
&=GR^{(3,1)}\!\left[
X_3^{(1)}\ket{\Gamma_2}
\otimes
H_3\ket{0}_3
\right]
\nonumber\\
&=\frac{1}{\sqrt{6}}\Bigl(
\ket{210}+\ket{011}+\ket{112}
-\ket{100}-\ket{201}-\ket{002}
\Bigr).
\label{gamma3(1)}
\end{align}

\begin{figure}[htb]
\centering
\includegraphics[width=0.42\linewidth,trim=6 6 6 6,clip]{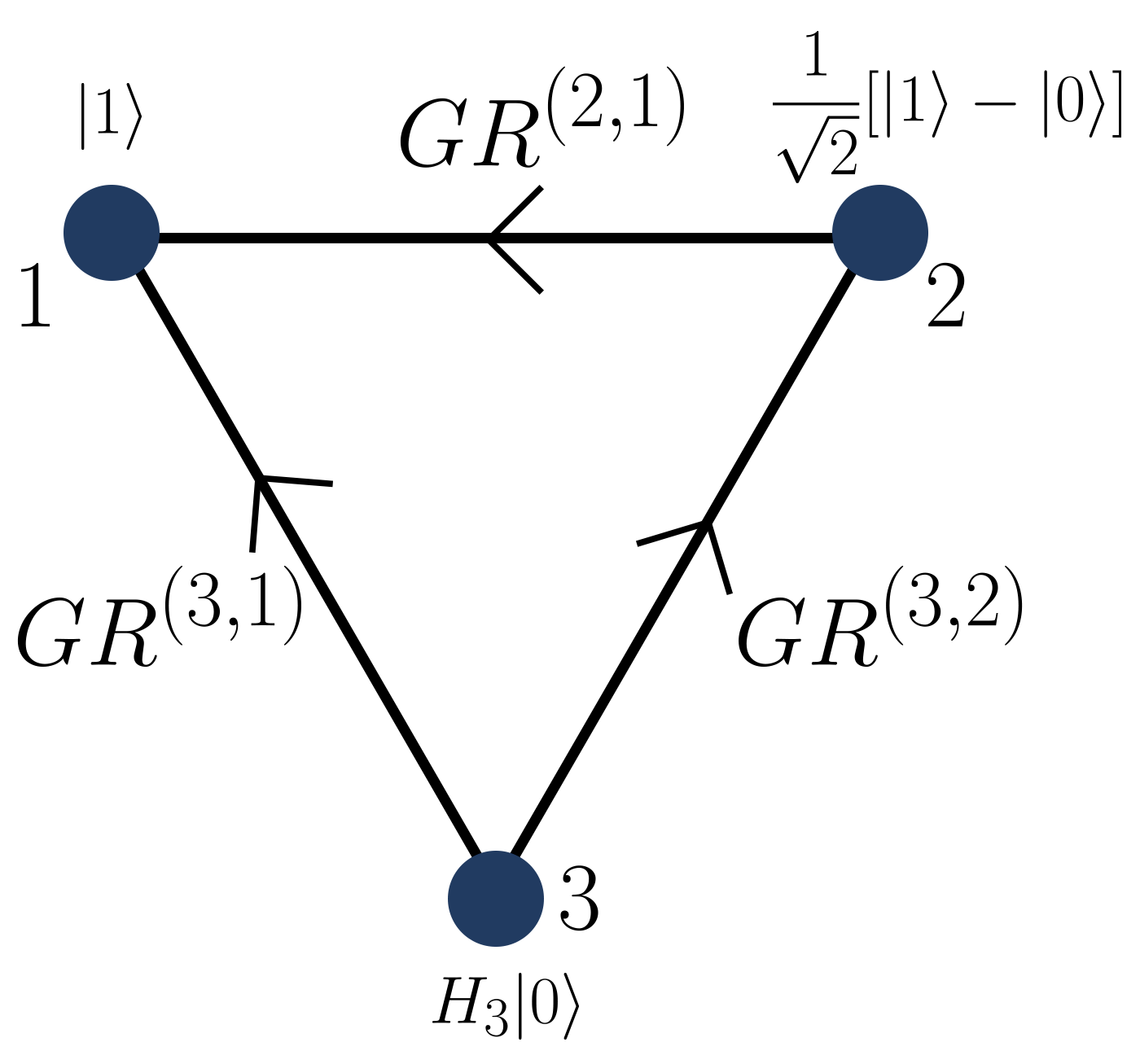}
\caption{Directed complete graph on three vertices whose orientation generates the fully antisymmetric state of three $d=3$ qudits. The directed edges represent the ordered application of the gates $GR^{(3,1)}$ and $GR^{(3,2)}$, leading to the state $\ket{\Gamma_3}=\ket{A_3}$.}
\label{grafo5}
\end{figure}

This state does not exhibit any particular exchange symmetry. We now add the second edge $3\to2$, as shown in Figure~\ref{grafo5}.

The resulting state is
\begin{align}
\ket{\Gamma_3}
&=
GR^{(3,1)}GR^{(3,2)}\!\left[
(X_3^{(1)}\otimes X_3^{(2)})\ket{\Gamma_2}
\otimes H_3\ket{0}_3 \right]
\nonumber\\
&=
\frac{1}{\sqrt{6}}\Bigl(
\ket{210}+\ket{021}+\ket{102}
-\ket{120}-\ket{201}-\ket{012}\Bigr),
\label{gamma3}
\end{align}
which is fully antisymmetric under particle permutations.

The state in Equation \eqref{gamma3} coincides exactly with $\ket{A_3}$ obtained from the general recursive construction in Equation \eqref{antis}. This shows that a complete directed graph on three vertices, endowed with a specific hierarchical orientation from the highest-indexed vertex to the lowest, reproduces the fully antisymmetric three-qudit state. {In Appendix~\ref{appk3} we illustrate, with the complete graph $K_3$, the fact that if the proper orientation is not respected, then the resulting graph state exhibits no symmetry. Additional explicit constructions for larger complete graphs are presented in Appendix~\ref{appk4}, where we detail the recursive generation of the states associated with $K_4$ and $K_5$.}

\section{Discussion and Future Perspectives}

In this work, we introduced a graph-based framework for the systematic construction of fully antisymmetric multipartite states, extending the standard notion of graph states beyond the symmetric setting generated by controlled-$Z$ interactions. In this sense, the $GR$-based construction provides a genuine generalization of graph states, encompassing new symmetry classes and requiring a richer class of graph structures. 

A central outcome of our analysis is the explicit connection between exchange symmetry and graph structure. In conventional graph states generated by $CZ$ gates, permutation symmetry is entirely determined by graph completeness and relies on the undirected nature and mutual commutativity of the interactions. By contrast, in graph states generated by the $GR$ operator, antisymmetry emerges only when the underlying graph is both complete and appropriately oriented. The directed character of the graph, reflecting the non-commutativity of the $GR$ gates, plays an essential role and has no counterpart in the standard graph-state formalism.

Together with the complete proof presented in the {\it Full Antisymmetry of the State for an Arbitrary Number of Qudits Section,} these results establish a precise structural correspondence: complete undirected graphs generate fully symmetric states, while complete directed graphs endowed with a hierarchical orientation generate fully antisymmetric states. This correspondence highlights orientation as a key resource for encoding fermionic exchange statistics within a graph-theoretic language. It is important to note that graph states generated by the $GR$ operator are not restricted to fully antisymmetric states, but may also belong to other classes of entangled states. For a fixed edge set, different choices of edge orientations lead to distinct quantum states, while preserving the underlying adjacency structure of the graph.

Beyond its conceptual implications, the present framework opens several directions for future research. From the perspective of quantum information theory, it raises the possibility of extending measurement-based and graph-based protocols to settings involving antisymmetric or mixed-symmetry states. From a structural standpoint, an important open question concerns the stabilizer properties of $GR$-generated graph states and their relation to existing stabilizer formalisms. More broadly, the graph-based encoding of exchange symmetry introduced here may provide a useful tool for exploring fermionic networks and many-body quantum systems within a unified and transparent graphical framework.

\section*{Acknowledgements}

M.R.d.J. and R.M.A. thank the Brazilian funding agency CNPq under grant codes PIBIC-UFPR/CNPq and 305957/2023-6, respectively. E.H. thanks the USIAS (University of Strasbourg Institute for Advanced Studies) for its partial support.

\appendix
\section{Inversion of the Orientation of the \boldmath{$K_3$} Graph}
\label{appk3}

\begin{figure}[htb]
\centering
\includegraphics[width=0.42\linewidth,trim=6 6 6 6,clip]{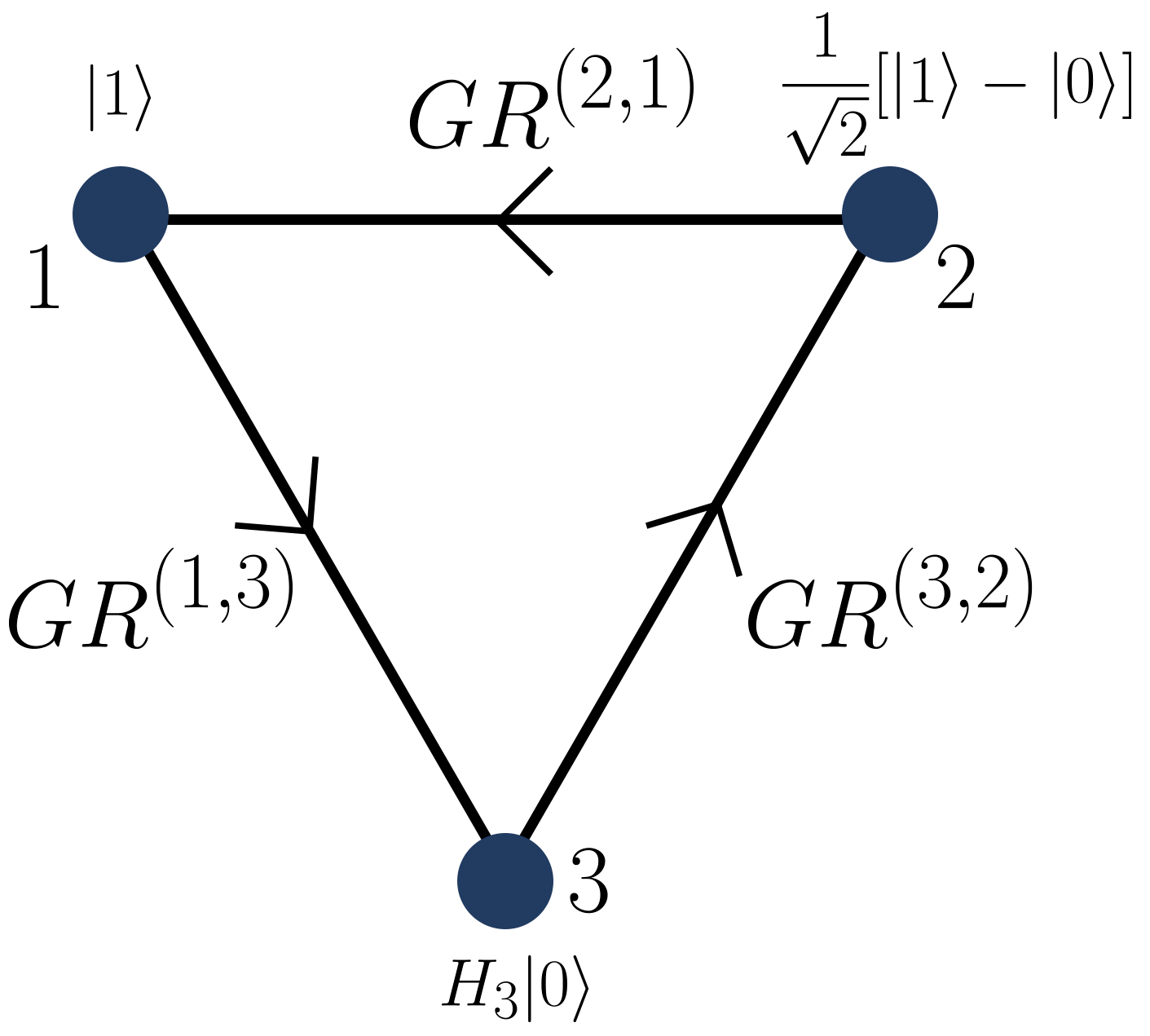}
\caption{Directed complete graph on three vertices whose orientation does not produce an antisymmetric state.}
\label{IK3}
\end{figure}

To illustrate that the orientation of the graph plays a key role in generating antisymmetric graph states, we invert the orientation of a single edge in the graph previously shown in Figure~\ref{grafo5} and calculate the resulting graph state. The modified graph is presented in Figure~\ref{IK3}. Recall that the notation $K_n$ denotes a complete graph, meaning that every pair of distinct vertices is connected by an edge. This notation does not imply any specific orientation of the graph.

Calculating the corresponding graph state, we obtain
\begin{align}
|\Gamma_3\rangle &= \frac{1}{\sqrt{6}} GR^{(1,3)} GR^{(3,2)} \Big[ \big(|12\rangle - |21\rangle\big) \otimes \big(|0\rangle + |1\rangle + |2\rangle\big) \Big] \nonumber\\
&= \frac{1}{\sqrt{6}} GR^{(1,3)} GR^{(3,2)} \Big( |120\rangle - |210\rangle + |121\rangle - |211\rangle + |122\rangle - |212\rangle \Big)\nonumber \\
&= \frac{1}{\sqrt{6}} GR^{(1,3)} \Big( |1 {1}0\rangle - |2 {2}0\rangle + |121\rangle - |2 {0}1\rangle + |1 {0}2\rangle - |212\rangle \Big) \nonumber\\
&= \frac{1}{\sqrt{6}} \Big( |11 {1}\rangle - |22 {2}\rangle + |12 {0}\rangle - |201\rangle + |102\rangle - |21 {0}\rangle \Big).
\label{gamma3invert}
\end{align}
The state in \eqref{gamma3invert} exhibits no permutation symmetry. This illustrates that different edge orientations generate distinct graph states and that antisymmetric states arise only for the specific orientation described previously, highlighting the decisive role played by edge orientation in determining the symmetry properties of the resulting graph state.

\section{Calculation of the \boldmath{$GR$} Graph States Associated with \boldmath{$K_4$}  and \boldmath{$K_5$}}
\label{appk4}

In this appendix, we explicitly construct the antisymmetric states for four and five qudits. These constructions illustrate the recursive graph-state algorithm introduced in the main text and demonstrate how it generates the antisymmetric states as the system size increases. Using the definition \eqref{Eqjesus2}, we obtain
\begin{equation}
    \ket{\Gamma_4}=GR^{(4,1)}GR^{(4,2)}GR^{(4,3)}(\tilde{X}_4\ket{\Gamma_3}\otimes H_4\ket{0}_4).
\end{equation}
Expanding the shift acting on the state $\ket{\Gamma_3}$, we find
\begin{align}
|\Gamma_4\rangle &= \prod_{i=1}^{3} GR^{(4,i)} \left( \frac{1}{\sqrt{3!}} (|321\rangle + |132\rangle + |213\rangle - |231\rangle - |312\rangle - |123\rangle) \otimes H_4 |0\rangle_4 \right) \nonumber\\
&= GR^{(4,1)} GR^{(4,2)} GR^{(4,3)} \frac{1}{\sqrt{4!}} \left( (|321\rangle + |132\rangle + \dots - |123\rangle) \otimes \sum_{k=0}^{3} (-1)^{3k} |k\rangle_4 \right).
\end{align}
Taking the tensor product and expanding the sum, we have
\begin{align}
|\Gamma_4\rangle &= \frac{1}{\sqrt{4!}} GR^{(4,1)} GR^{(4,2)} GR^{(4,3)} \Big( \nonumber\\
&\quad + |3210\rangle - |3211\rangle + |3212\rangle - |3213\rangle \nonumber\\
&\quad + |1320\rangle - |1321\rangle + |1322\rangle - |1323\rangle \nonumber\\
&\quad + |2130\rangle - |2131\rangle + |2132\rangle - |2133\rangle \nonumber\\
&\quad - |2310\rangle + |2311\rangle - |2312\rangle + |2313\rangle \nonumber\\
&\quad - |3120\rangle + |3121\rangle - |3122\rangle + |3123\rangle \nonumber\\
&\quad - |1230\rangle + |1231\rangle - |1232\rangle + |1233\rangle \Big).
\end{align}
Applying $GR^{(4,3)}$, we obtain
\begin{align}
|\Gamma_4\rangle &= \frac{1}{\sqrt{4!}} GR^{(4,1)} GR^{(4,2)} \Big( \nonumber\\
&\quad + |3230\rangle - |3201\rangle + |3212\rangle - |3223\rangle \nonumber\\
&\quad + |1320\rangle - |1331\rangle + |1302\rangle - |1313\rangle \nonumber\\
&\quad + |2110\rangle - |2121\rangle + |2132\rangle - |2103\rangle \nonumber\\
&\quad - |2330\rangle + |2301\rangle - |2312\rangle + |2323\rangle \nonumber\\
&\quad - |3120\rangle + |3131\rangle - |3102\rangle + |3113\rangle \nonumber\\
&\quad - |1210\rangle + |1221\rangle - |1232\rangle + |1203\rangle \Big).
\end{align}
Applying $GR^{(4,2)}$, we find
\begin{align}
|\Gamma_4\rangle &= \frac{1}{\sqrt{4!}} GR^{(4,1)} \Big( \nonumber\\
&\quad + |3230\rangle - |3301\rangle + |3012\rangle - |3123\rangle \nonumber\\
&\quad + |1120\rangle - |1231\rangle + |1302\rangle - |1013\rangle \nonumber\\
&\quad + |2310\rangle - |2021\rangle + |2132\rangle - |2203\rangle \nonumber\\
&\quad - |2130\rangle + |2201\rangle - |2312\rangle + |2023\rangle \nonumber\\
&\quad - |3320\rangle + |3031\rangle - |3102\rangle + |3213\rangle \nonumber\\
&\quad - |1210\rangle + |1321\rangle - |1032\rangle + |1103\rangle \Big).
\end{align}
Finally, applying $GR^{(4,1)}$, we obtain the antisymmetric graph state for four vertices
\begin{align}
|\Gamma_4\rangle &= \frac{1}{\sqrt{4!}} \Big( \nonumber\\
&\quad + |1230\rangle - |2301\rangle + |3012\rangle - |0123\rangle \nonumber \\
&\quad + |3120\rangle - |0231\rangle + |1302\rangle - |2013\rangle \nonumber\\
&\quad + |2310\rangle - |3021\rangle + |0132\rangle - |1203\rangle \nonumber\\
&\quad - |2130\rangle + |3201\rangle - |0312\rangle + |1023\rangle \nonumber\\
&\quad - |1320\rangle + |2031\rangle - |3102\rangle + |0213\rangle \nonumber\\
&\quad - |3210\rangle + |0321\rangle - |1032\rangle + |2103\rangle \Big).
\end{align}

Using the same method to calculate the complete graph state for five qudits, we obtain the following result:
\begin{align}
|\Gamma_5\rangle = &\frac{1}{\sqrt{5!}}\big(
|23140\rangle + |34201\rangle + |40312\rangle + |01423\rangle + |12034\rangle \nonumber\\
&+ |31240\rangle + |42301\rangle + |03412\rangle + |14023\rangle + |20134\rangle \nonumber\\
&+ |12340\rangle + |23401\rangle + |34012\rangle + |40123\rangle + |01234\rangle \nonumber\\
&- |32140\rangle - |43201\rangle - |04312\rangle - |10423\rangle - |21034\rangle \nonumber\\
&- |13240\rangle - |24301\rangle - |30412\rangle - |41023\rangle - |02134\rangle \nonumber\\
&- |21340\rangle - |32401\rangle - |43012\rangle - |04123\rangle - |10234\rangle \nonumber\\
&- |12430\rangle - |23041\rangle - |34102\rangle - |40213\rangle - |01324\rangle \nonumber\\
&- |24130\rangle - |30241\rangle - |41302\rangle - |02413\rangle - |13024\rangle \nonumber\\
&- |41230\rangle - |02341\rangle - |13402\rangle - |24013\rangle - |30124\rangle \nonumber\\
&+ |21430\rangle + |32041\rangle + |43102\rangle + |04213\rangle + |10324\rangle \nonumber\\
&+ |14230\rangle + |20341\rangle + |31402\rangle + |42013\rangle + |03124\rangle \nonumber\\
&+ |42130\rangle + |03241\rangle + |14302\rangle + |20413\rangle + |31024\rangle \nonumber\\
&+ |41320\rangle + |02431\rangle + |13042\rangle + |24103\rangle + |30214\rangle \nonumber\\
&+ |13420\rangle + |24031\rangle + |30142\rangle + |41203\rangle + |02314\rangle \nonumber\\
&+ |34120\rangle + |40231\rangle + |01342\rangle + |12403\rangle + |23014\rangle \nonumber\\
&- |14320\rangle - |20431\rangle - |31042\rangle - |42103\rangle - |03214\rangle \nonumber\\
&- |31420\rangle - |42031\rangle - |03142\rangle - |14203\rangle - |20314\rangle \nonumber\\
&- |43120\rangle - |04231\rangle - |10342\rangle - |21403\rangle - |32014\rangle \nonumber\\
&- |34210\rangle - |40321\rangle - |01432\rangle - |12043\rangle - |23104\rangle \nonumber\\
&- |42310\rangle - |03421\rangle - |14032\rangle - |20143\rangle - |31204\rangle \nonumber\\
&- |23410\rangle - |34021\rangle - |40132\rangle - |01243\rangle - |12304\rangle \nonumber\\
&+ |43210\rangle + |04321\rangle + |10432\rangle + |21043\rangle + |32104\rangle \nonumber\\
&+ |24310\rangle + |30421\rangle + |41032\rangle + |02143\rangle + |13204\rangle \nonumber\\
&+ |32410\rangle + |43021\rangle + |04132\rangle + |10243\rangle + |21304\rangle\big).
\end{align}

The purpose of this appendix is to illustrate the recursive graph-state construction introduced in the main text through the explicit cases $n=4$ and $n=5$. For larger systems, the number of terms in the resulting superposition grows factorially, making direct analytical expansions increasingly cumbersome, so computational methods are more appropriate.

\bibliographystyle{apsrev4-2}
\bibliography{references}

\end{document}